\documentclass[12pt]{article}
\usepackage{graphicx}
\def\Journal#1#2#3#4{{#1} {\bf #2}, #3 (#4)}

\def\NPB{Nucl. Phys. B}
\def\PLB{Phys. Lett. B}

\def\PRD{Phys. Rev. D}

\def\EPJ{Eur. Phys. J. C}

\newcommand{\ra}{\rightarrow}
\newcommand{\Dbar}{\bar{D}}
\def\lsim{\mathrel{\rlap{\lower4pt\hbox{\hskip1pt$\sim$}}
    \raise1pt\hbox{$<$}}}

\begin{document}
\thispagestyle{empty}

\vspace*{1cm}
\begin{center}
\baselineskip=1cm
{\Large \bf $D$ Meson Production in Neutrino DIS 
and Polarized Strange Quark Distribution} 

\baselineskip=0.6cm
\vspace{1.5cm}
Kazutaka ~SUDOH \\
\vspace{0.5cm}
{\em Radiation Laboratory, \\
RIKEN (The Institute of Physical and Chemical Research), \\
Wako, Saitama 351-0198, JAPAN} \\
\vspace{0.2cm}
E-mail: {\tt sudou@rarfaxp.riken.go.jp}
\end{center}

\vspace{3cm} 
\noindent
\begin{center}
{\bf Abstract}
\end{center}

Semi-inclusive $D$/$\Dbar$ meson productions in neutrino deep inelastic 
scattering are studied including ${\cal O}(\alpha_s)$ corrections.
Supposing a future neutrino factory, cross sections and spin asymmetries 
in polarized processes are calculated by using various parametrization models 
of polarized parton distribution functions.
We suggest that $\Dbar$ production is promising to directly extract 
the strange quark distribution. 

\vspace{0.5cm}
\clearpage

\setcounter{page}{1}
\section{Introduction}
In recent years, experimental data for heavy flavor (charm or bottom quark) 
production and decay are reported by several collaborations.
Since the heavy quark mass scale is quite larger than $\Lambda_{QCD}$, it is 
considered that we can treat heavy quarks purely perturbatively.
Heavy quarks are produced only at the short distance scale within the 
framework of a fixed flavor number scheme (FFNS), where only light quarks 
($u$, $d$, and $s$) and gluons are considered as active partons, and any heavy 
quark ($c$, $b$, ...) contributions are calculated in fixed order $\alpha_s$ 
perturbation theory.
Physics of heavy flavor is relatively undisturbed by non-perturbative effects.

A study of heavy flavor production in deep inelastic scattering (DIS) is 
one of the most promising ways to access the parton density in the nucleon.
As is well known, the polarized parton distribution function (PDF) plays 
an important role in deep understandings of spin structure of the nucleon.
In particular, flavor structure of sea quark distributions has been actively 
studied in these years.
Most of parametrization models are so far assumed the flavor SU(3)$_f$ 
symmetry to determine the sea quark distributions.
However, there is an attempt to include the violation effects of the SU(3)$_f$ 
symmetrty \cite{Leader99}. 
We have expected that the sea quark flavor structure is investigated by 
analysing the semi-inclusive DIS data in more details, whereas only the 
combinations $\Delta q(x, Q^2)+\Delta \bar{q}(x, Q^2)$ can be determined in 
inclusive DIS.
All analyses of the traditional inclusive DIS data suggest that the polarized 
strange quark distribution $\Delta s(x, Q^2)+\Delta \bar{s}(x, Q^2)$ is 
significantly negative \cite{Leader02}.
However, more recently the HERMES collaboration has reported preliminary 
results about the polarized strange quark distribution \cite{stosslein02}, 
in which $\Delta s(x)+\Delta \bar{s}(x)$ at $Q^2 =2.5$ GeV$^2$ is slightly 
positive.
Thus, knowledge about the polarized sea quark distributions remain still poor, 
and theoretical and experimental ambiguities are rather large. 
In order to understand the spin and flavor structure of the nucleon, we need 
more information about the polarized strange quark distribution functions.

Charged current (CC) DIS is effective to extract the flavor decomposed 
polarized PDFs, since $W^{\pm}$ boson changes the flavor of parton.
Since there is no intrinsic heavy flavor component in the FFNS, 
we can extract information about the parton flavor in the nucleon from 
the study of heavy flavor production in CC DIS.
Actually, the NuTeV collaboration reported a measurement of unpolarized 
$s$ and $\bar{s}$ quark distributions by measuring dimuon cross sections 
in neutrino-DIS \cite{NuTeV01}.
Neutrino-induced CC DIS is interesting and challenging as not only the 
unpolarized reaction but also the polarized reaction \cite{Forte01} at a 
future neutrino factory \cite{Mangano01}.

\begin{figure}[t!]
\hspace*{2cm}
(a)
\hspace*{1cm}
\begin{minipage}{5cm}
   \includegraphics[scale=0.7,clip]{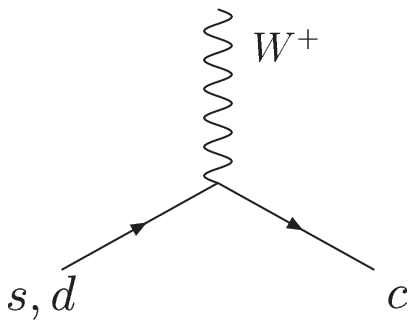}\\
\end{minipage}
\\
\hspace*{2cm}
(b)
\hspace*{1cm}
\begin{minipage}{7cm}
   \includegraphics[scale=0.7,clip]{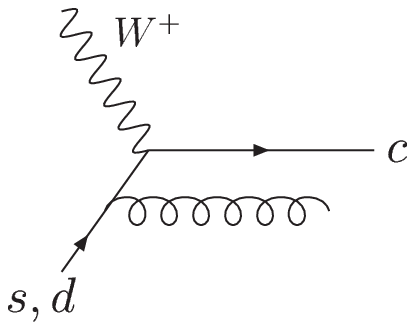}
   \includegraphics[scale=0.7,clip]{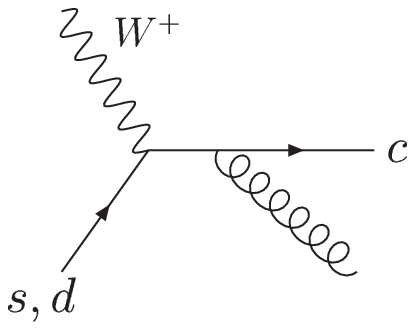}\\
\end{minipage}
\\
\hspace*{2cm}
(c)
\hspace*{1cm}
\begin{minipage}{10cm}
   \includegraphics[scale=0.7,clip]{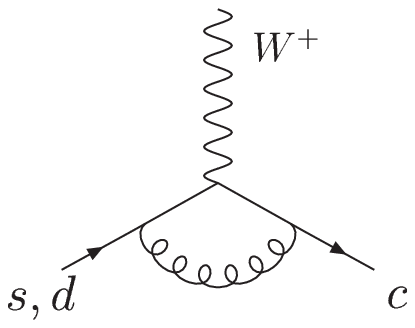}
   \includegraphics[scale=0.7,clip]{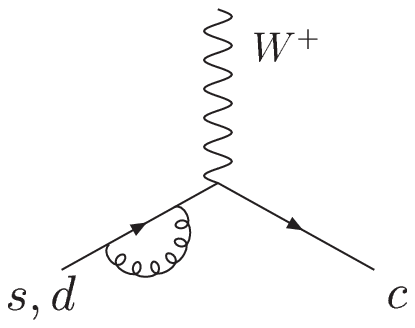}
   \includegraphics[scale=0.7,clip]{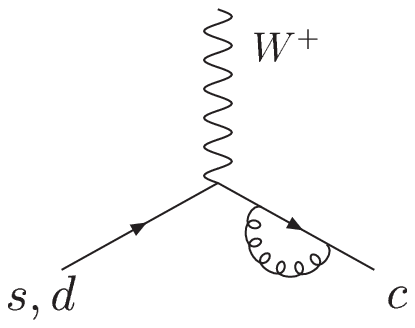}\\
\end{minipage}
\\
\hspace*{2cm}
(d)
\hspace*{1cm}
\begin{minipage}{7cm}
   \includegraphics[scale=0.7,clip]{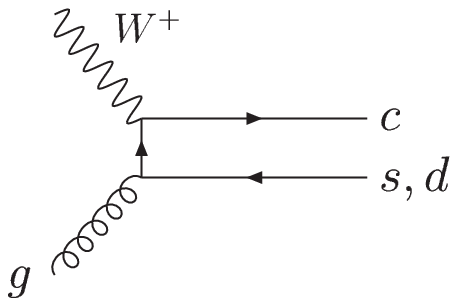}
   \includegraphics[scale=0.7,clip]{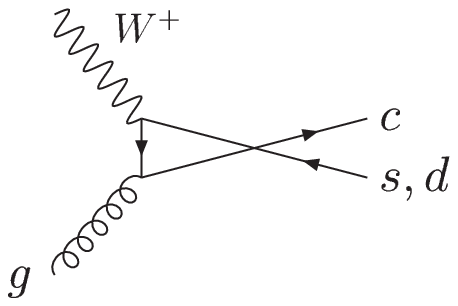}
\end{minipage}
\caption{Feynman diagrams of subprocesses for $D$ production in CC DIS 
up to ${\cal O}(\alpha_s)$: 
(a) Born term, (b) real gluon radiation, (c) virtual corrections, 
and (d) boson-gluon fusion processes. 
All quarks are replaced by anti-quarks in $\Dbar$ production case.
}
\label{diagram}
\end{figure}

In this work, to extract information about the polarized PDFs we investigated 
$D$/$\Dbar$ meson productions in CC DIS including ${\cal O}
(\alpha_{s})$ corrections in neutrino and polarized proton scattering: 
\begin{equation}
\nu + \vec{p} \ra l^- + D + X, 
\end{equation}
\begin{equation}
\bar{\nu} + \vec{p} \ra l^+ + \Dbar + X.
\end{equation}
Typical feynman diagrams of subprocesses are illustrated in 
Fig. \ref{diagram} up to ${\cal O}(\alpha_{s})$ next-to-leading order (NLO) 
corrections.
The leading order process is due to $W^{\pm}$ boson exchange $W^+ s(d)\ra c$ 
(Fig. \ref{diagram} (a)).
In addition, several processes are taken account in NLO calculations, 
in which (b) real gluon radiation processes $W^+ s(d) \ra cg$, (c) virtual 
corrections to remove the singularity coming from soft gluon radiation, and 
(d) boson-gluon fusion processes $W^+ g\ra c\bar{s}(\bar{d})$ are considered.
Thus, the process is quite sensitive to the strange quark and gluon 
distribution functions in the nucleon.
These processes might be observed in the forthcoming neutrino experiments, 
though there is no experiment using neutrino beams with the polarized 
target at present.

\section{Charm Production in CC DIS}
We have calculated the spin-independent and -dependent cross 
sections, and the spin asymmetry $A^D$ which is defined by
\begin{equation}
A^D \equiv \frac{[d\sigma_{--} - d\sigma_{-+}]/dx}
{[d\sigma_{--} + d\sigma_{-+}]/dx}
=\frac{d\Delta\sigma /dx}{d\sigma /dx} ,
\label{asymmetry}
\end{equation}
where $d\sigma_{hh'}$ denotes the spin-dependent cross section with definite 
helicities $h$ and $h'$ for neutrino and the target proton, respectively.
Since neutrino is naturally polarized, neutrino and anti-neutrino beams have 
only helicity state $-$ and $+$, respectively.
Therefore, for $\Dbar$ production case, the spin asymmetry $A^{\Dbar}$ is 
obtained by replacing the neutrino beam helicity $-$ to $+$ 
in Eq. (\ref{asymmetry}).
The spin-dependent differential cross section for $\nu \vec{p}$ scattering can 
be written in terms of the polarized structure functions $g_i$ as follows:
\begin{equation}
\frac{d^{2[3]} \Delta\sigma^{\nu p}}{dxdy[dz]}=\frac{G_F^2 s}{2\pi 
(1+Q^2/M_W^2)^2}
\left[ (1-y) g_4^{W^{\mp}} + y^2 x g_3^{W^{\mp}} 
\pm y(1-\frac{y}{2})x g_1^{W^{\mp}} \right] ,
\end{equation}
where $Q^2 = -q^2$ and $G_F$, $s$, and $M_W$ denote the Fermi coupling, 
center of mass energy squared, and $W^{\pm}$ boson mass, respectively.
Note that the $+$ and $-$ in front of the 3rd term correspond to 
when initial beam is anti-neutrino and neutrino, respectively.
Kinematical variables $x$ and $y$ are Bjorken scaling variable and 
inelasticity defined according to the standard DIS kinematics, and $z$ is 
defined by $z=P_p \cdot P_D/P_p \cdot q$ with $P_p$, $P_D$ and $q$ being the 
momentum of proton, $D$ meson, and $W^{\pm}$ boson, respectively.
The polarized structure functions $g_i$ in $\nu \vec{p}$ scattering are 
obtained by the following convolutions:
\begin{eqnarray}
{\cal G}_i (x,z,Q^2) &=& \Delta s' (\xi,\mu^2_F)D_c(z) \nonumber \\
&&+
\frac{\alpha_s(\mu^2_R)}{2\pi}
\int_{\xi}^1 \frac{d\xi '}{\xi '} 
\int_{\max(z,\zeta_{\min})}^1 \frac{d\zeta}{\zeta} 
\left\{ \Delta H_i^q (\xi ',\zeta,\mu^2_F,\lambda) 
\Delta s' (\frac{\xi}{\xi '},\mu^2_F) \right. \nonumber \\
&&+
\left. \Delta H_i^g (\xi ',\zeta,\mu^2_F,\lambda) 
\Delta g(\frac{\xi}{\xi '},\mu^2_F)
\right\}
D_c(\frac{z}{\zeta}) ,
\end{eqnarray}
where $\Delta s'$ means $\Delta s' \equiv |V_{cs}|^2 \Delta s + 
|V_{cd}|^2 \Delta d$ with CKM parameters.
$\Delta H_i^{q, g}$ are coefficient functions of quarks and gluons, which can 
be calculated by using perturbative QCD.
The argument $\xi$ is the slow rescaling parameter, and $\xi^{\prime}$ and 
$\zeta$ are the partonic scaling variables which are defined for the parton 
momentum $p_i$ as
\begin{equation}
\xi=\frac{Q^2}{2P_p \cdot q}\left(1+\frac{m_c^2}{Q^2}\right), ~~
\xi'=\frac{Q^2}{2p_{s, g} \cdot q}\left(1+\frac{m_c^2}{Q^2}\right), ~~
\zeta=\frac{p_{s, g}\cdot p_c}{p_{s, g}\cdot q} ,
\end{equation}
where $p_g$, $p_s$, $p_c$, and $m_c$ are the momentum of gluon, $s$ quark, 
$c$ quark, and the charm quark mass, respectively.
$D_c (z)$ represents the fragmentation function of an outgoing charm quark 
decaying to $D$ meson.
For the fragmentation function, we adopted the parametrization proposed by 
Peterson {\it et al.} \cite{Peterson83} and recently developed by Kretzer 
{\it et al.} \cite{Kretzer99}.
${\cal G}_i (x,z,Q^2)$ is related to the polarized structure functions through 
${\cal G}_1 \equiv g_1 /2$, ${\cal G}_3 \equiv g_3$, and 
${\cal G}_4 \equiv g_4 /2\xi$.
Similar analyses have been done by Kretzer {\it et al.}, 
in which charged current charm production at NLO in $ep$ \cite{Kretzer99-2} 
and $\nu p$ \cite{Kretzer97} scattering is discussed.

\section{Numerical Results}
In numerical calculations, we set a charm quark mass $m_c =1.4$ GeV, 
an initial neutrino beam energy $E_{\nu}=200$ GeV, and 
the factorization scale $\mu_F$ which is equal to the renormalization 
scale $\mu_R$ as $\mu_F^2 = \mu_R^2 = Q^2 +m_c^2$.
We used the GRV98 \cite{GRV98} and MRST99 \cite{MRST99} parametrizations as 
the unpolarized PDFs.
As for the polarized PDFs, we adopted the following four kinds of 
parametrizations; AAC00 \cite{AAC00}, BB02 \cite{BB02}, GRSV01 \cite{GRSV01}, 
and LSS02 \cite{LSS02}, which are recently proposed and now widely used. 

\begin{figure}[t!]
   \includegraphics[scale=0.3,clip]{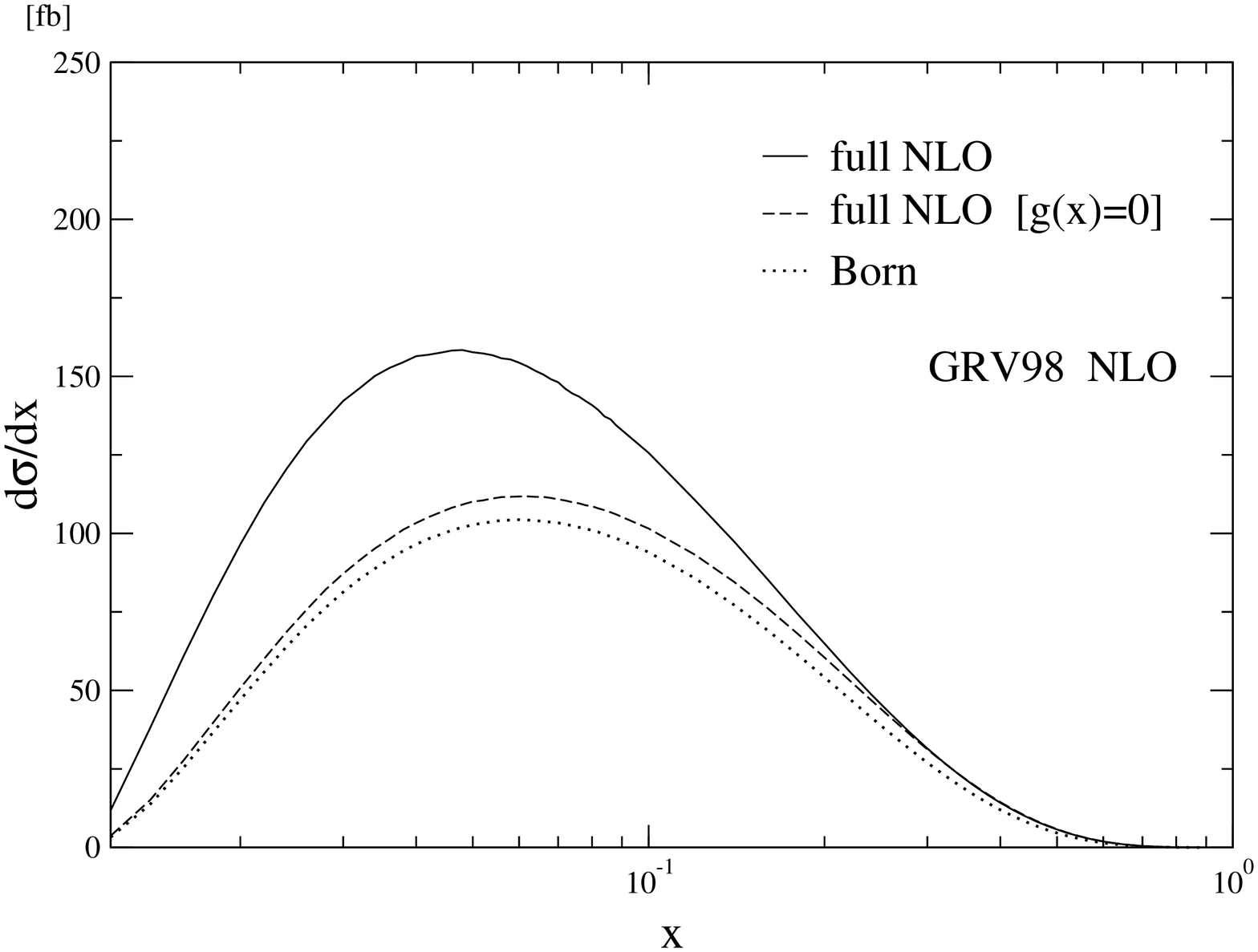}
\hspace{0.2cm}
   \includegraphics[scale=0.3,clip]{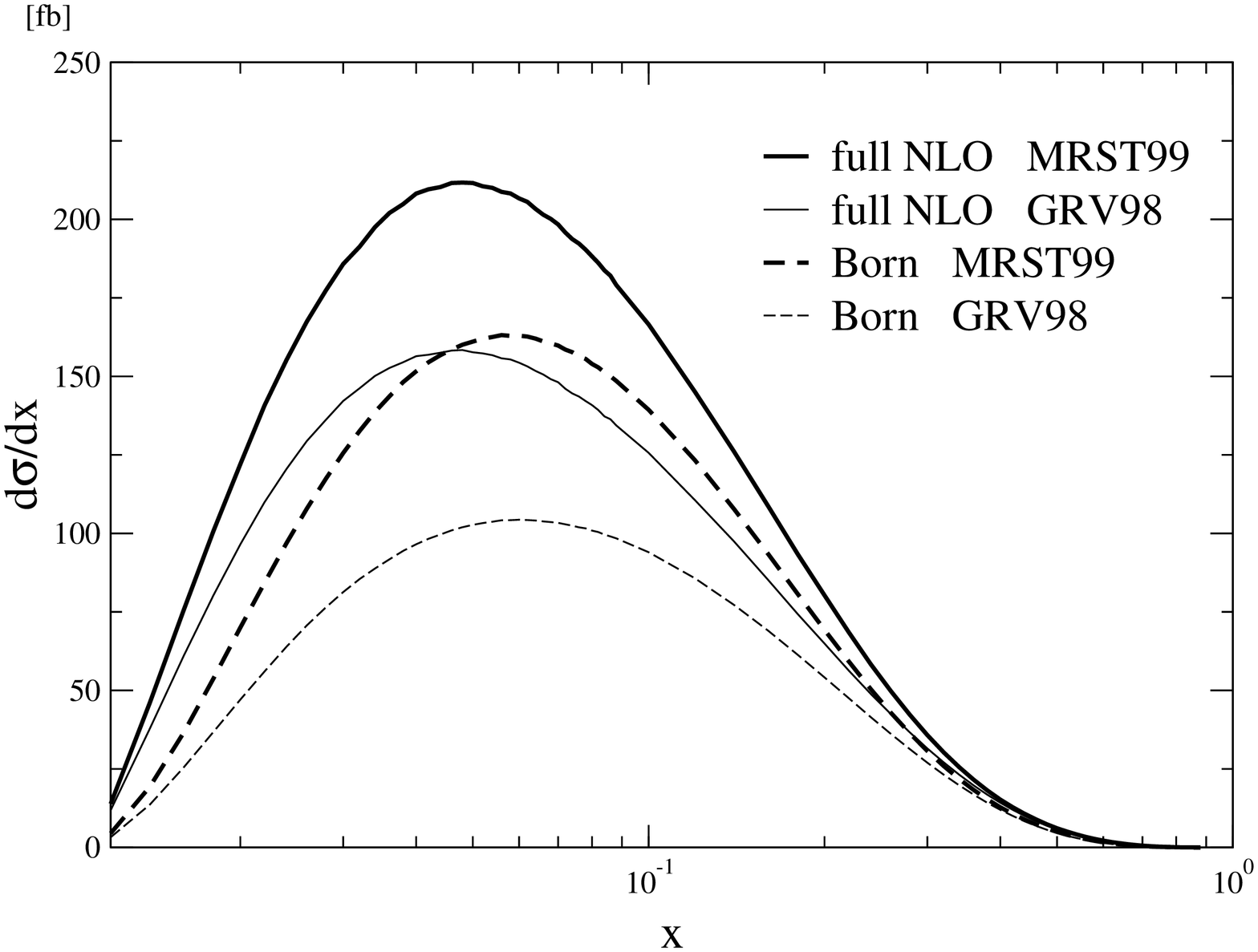}
\caption{The spin-independent differential cross sections for the process 
$\nu \vec{p} \ra l^- D X$ at $E_{\nu}=200$ GeV as a function of $x$.
We show the contribution from each diagram (left panel) and parametrization 
model dependence of the unpolarized PDFs (right panel).
Solid, dashed, and dotted lines in the left panel show the cross section in 
full NLO, full NLO with $g(x)=0$, and LO calculation, respectively.
Solid and dashed lines in the right panel represent the case of MRST99 and 
GRV98 parametrizations, and bold and normal lines are full NLO and LO cross 
sections, respectively.}
\label{fig-Crs-Unpol}
\end{figure}

We show the spin-independent differential cross section for $D$ meson 
production $\nu \vec{p} \ra l^- D X$ in Fig. \ref{fig-Crs-Unpol}.
The left panel in Fig. \ref{fig-Crs-Unpol} represents the contribution to the 
cross section from each diagram illustrated in Fig. \ref{diagram}.
Solid, dashed, and dotted lines indicate the contribution form full NLO, 
full NLO with $g(x)=0$, and LO diagrams to the $x$ differential 
cross section, respectively.
Hence, the difference between the dashed and dotted lines comes from the gluon 
radiation process $W^+ s(d) \ra cg$ and virtual corrections 
(Fig. \ref{diagram} (b) and (c)), while the 
difference between the solid and dashed lines comes from the boson-gluon 
fusion process $W^+ g\ra c\bar{s}(\bar{d})$ (Fig. \ref{diagram} (d)).
As shown in Fig. \ref{fig-Crs-Unpol}, the contribution from the NLO 
boson-gluon fusion process is considerably large.
This is because the gluon distribution is sufficiently larger than the 
strange quark distribution, though the short distance matrix element 
in NLO is suppressed by the strong coupling constant $\alpha_s$.

Comparison of the cross section using MRST99 and GRV98 parametrizations for 
the unpolarized PDFs is represented in the right panel in 
Fig. \ref{fig-Crs-Unpol}.
The gap between two dashed lines stems from the difference of behavior of 
the unpolarized strange quark distribution, since only the strange quark 
distribution contributes to the cross section at the LO level.
We found that the parametrization model dependence is quite large. 
It is indicated that even the unpolarized PDFs still has large ambiguity, 
in spite of the analyses of unpolarized processes are investigated 
for a long time.
Therefore, this reaction is effective to determine the unpolarized strange 
quark distribution, because the charm quark production in CC DIS using 
neutrino beams is sensitive to the strange density in the nucleon.

\begin{figure}[t!]
   \includegraphics[scale=0.3,clip]{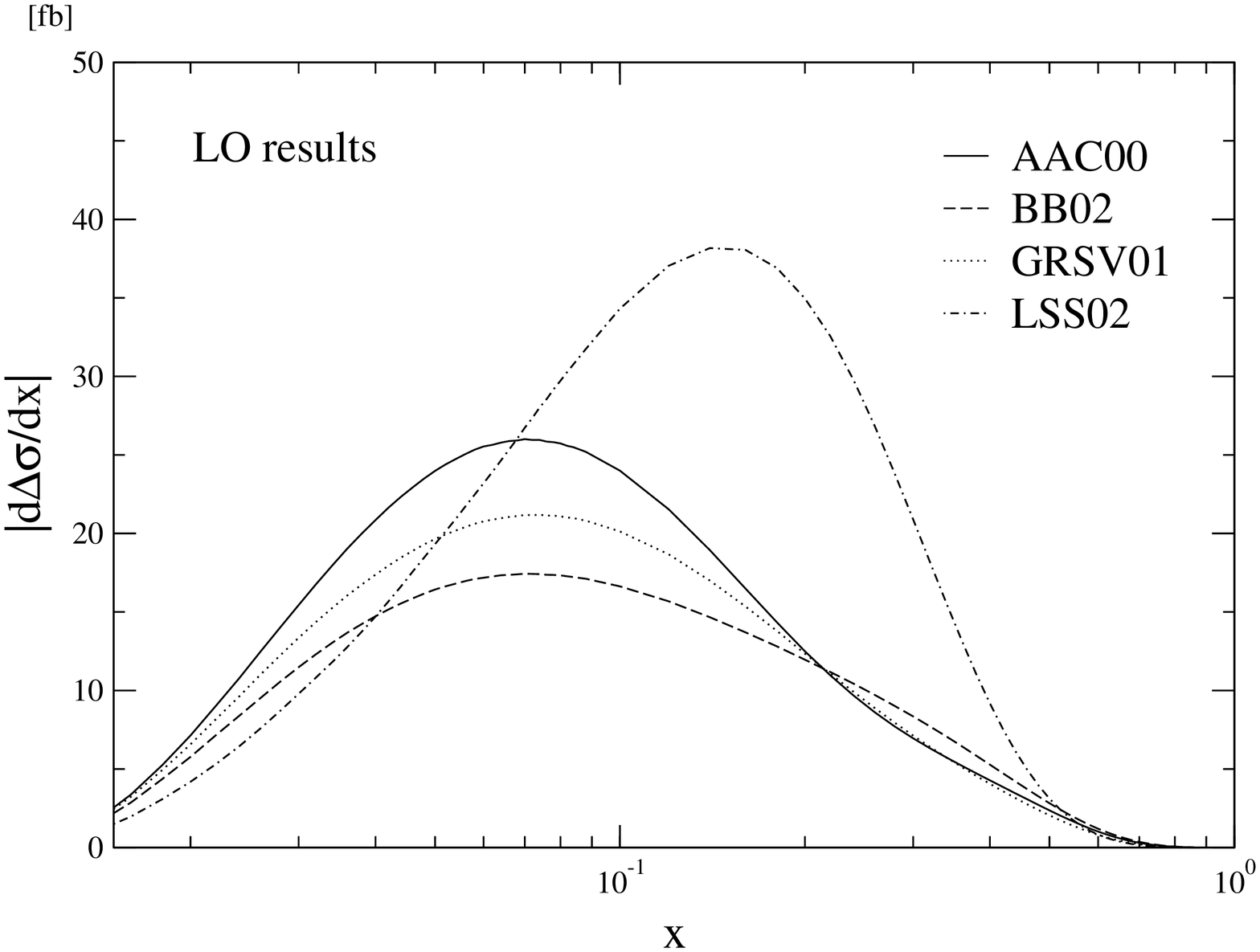}
\hspace{0.2cm}
   \includegraphics[scale=0.3,clip]{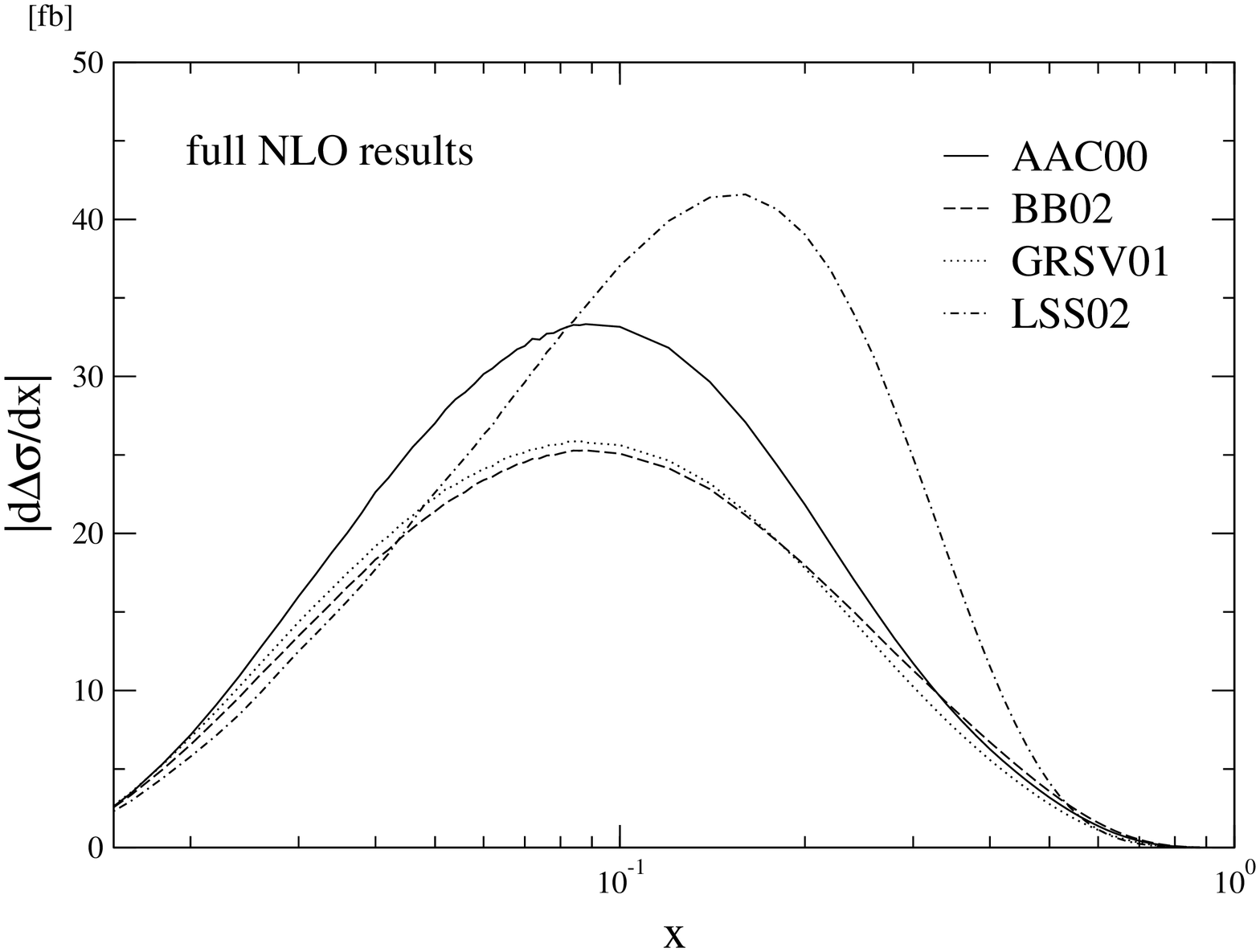}
\caption{The $x$ distribution of spin-dependent cross sections for $D$ 
meson production $\nu \vec{p} \ra l^- D X$ in LO (left panel) and full NLO 
(right panel).
Solid, dashed, dotted, and dot-dashed lines show the case of AAC00 set-2, 
BB02 scenario-2, GRSV01 standard set, and LSS02 $\overline{\mbox{MS}}$ 
parametrizations, respectively.}
\label{fig-Crs}
\end{figure}

The spin-dependent cross section for $D$ meson production is presented in 
Fig. \ref{fig-Crs}.
We show the comparison between LO (left panel) and full NLO (right panel)
results with various parametrization models of the polarized PDFs.
We see large contribution from NLO corrections and strong parametrization 
model dependence of the polarized PDFs.
As well as the unpolarized case, the cross sections are dominated by the LO 
process and boson-gluon fusion process at NLO. 
Contribution from the gluon radiation and virtual corrections 
are not significant in the cross sections.

Both the LO and full NLO results by the LSS parametrization are quite large 
compared with other parametrizations.
In the LSS parametrization, the polarized strange quark distribution has the 
peak at $x\sim 0.2$, whereas the polarized gluon distribution is not 
significant in this $x$ region.
Therefore, the LO cross section becomes large, and the difference between the 
LO and full NLO cross sections is consequently small for the LSS 
parametrization.
On the other hand, the NLO cross sections by GRSV and BB parametrizations are 
quite similar in whole $x$ regions, while we see some difference in the 
LO cross section.

\begin{figure}[t!]
\hspace*{-0.2cm}
   \includegraphics[scale=0.3,clip]{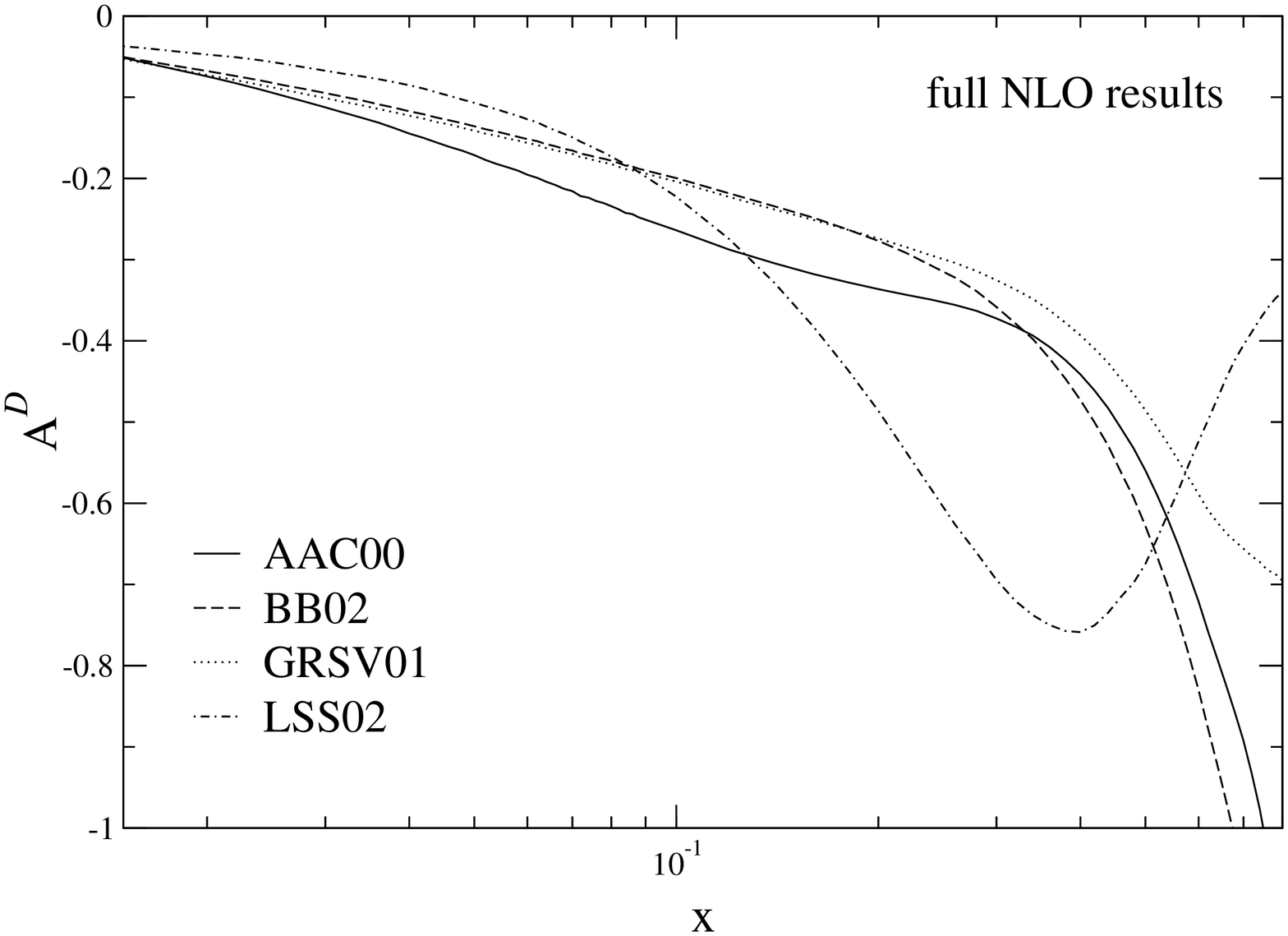}
\hspace{0.2cm}
   \includegraphics[scale=0.3,clip]{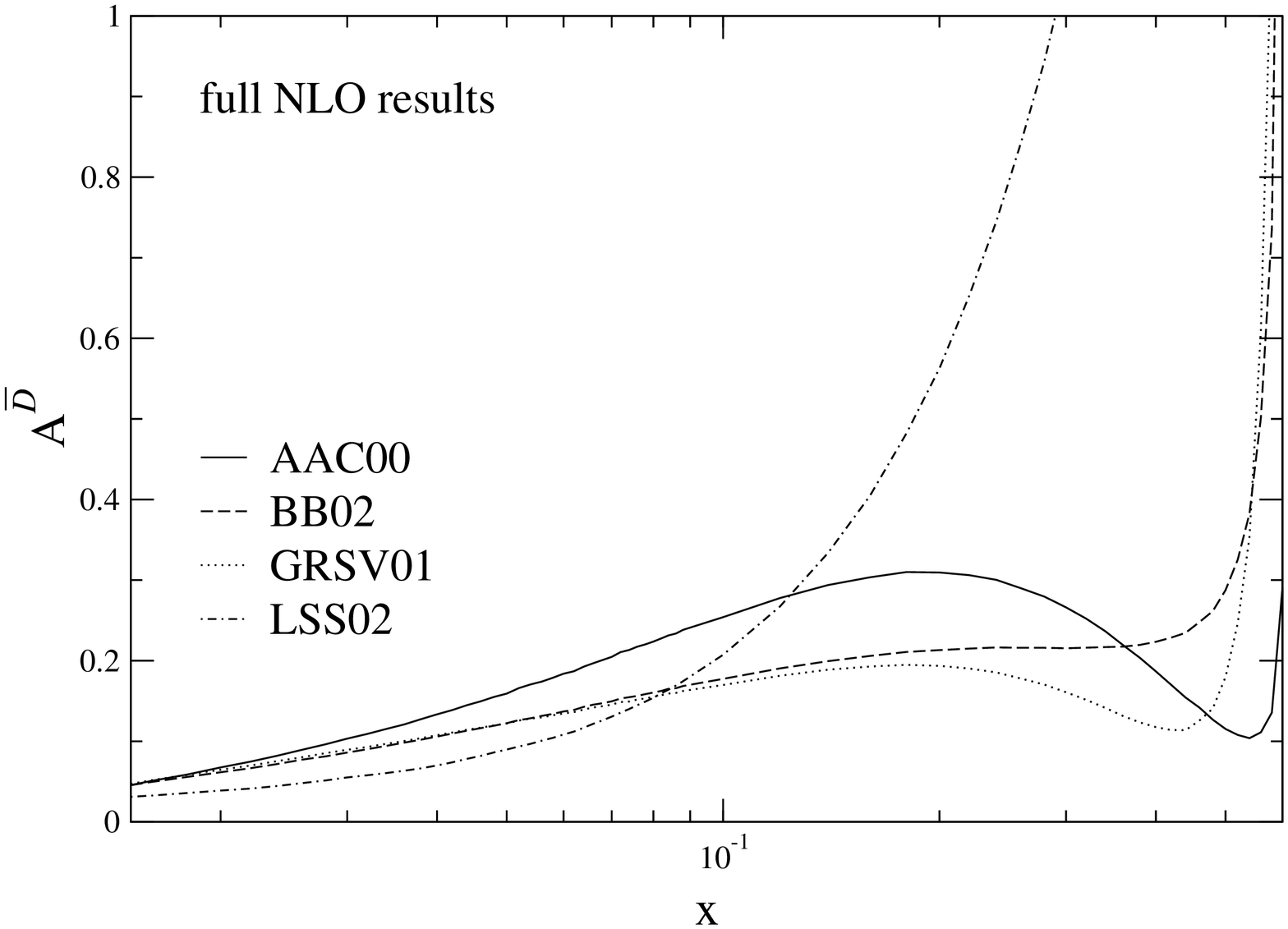}
\caption{Comparison of the spin asymmetries in NLO for $D$ production 
$\nu \vec{p} \ra l^- D X$ (left panel) and $\Dbar$ production 
$\bar{\nu} \vec{p} \ra l^+ \Dbar X$ (right panel) with various parametrization 
models of the polarized PDFs.
Several lines are the same as Fig. \ref{fig-Crs}.}
\label{fig-All}
\end{figure}

We show the spin asymmetry $A^D$ including ${\cal O}(\alpha_s)$ corrections 
in Fig. \ref{fig-All} as a function of $x$. 
Left panel and right panel in Fig. \ref{fig-All} represent asymmetries for 
$D$ production and $\Dbar$ production, respectively.
For $D$ production, $s$, $d$ quarks and gluon distribution contribute 
to the asymmetry $A^D$.
$A^D$ is dominated by the valence $d_v$ quark contribution at large $x$ 
regions ($x > 0.3$), though the $d$ quark component is quite highly 
suppressed by CKM.
On the contrary, for $\Dbar$ production, $\bar{s}$, $\bar{d}$ quarks 
and gluon component contribute to the asymmetry $A^{\Dbar}$.
The $\bar{d}$ quark contribution is almost negligible in $A^{\Dbar}$.
Therefore, the asymmetry is directly affected by the shape of the $\bar{s}$ 
quark distribution function.
Note that the difference of the sign between these two asymmetries comes 
from a definition of the asymmetry given in Eq. (\ref{asymmetry}).

As shown in both figures, spin asymmetries strongly depend on the 
parametrization models.  
We see that the case of the LSS parametrization is quite different from 
the ones of other parametrizations.
In particular, the asymmetry by the LSS parametrization in $\Dbar$ production 
goes over 1 at $x\sim 0.3$, though the asymmetry should be less than 1.
This is because the polarized strange quark distribution in their 
parametrization extremely violates the positivity condition at $x\sim 0.3$. 
Measurement of $\Dbar$ production in this reaction is effective 
to test the parametrization models of the polarized PDFs.
Also, we might directly measure the polarized strange quark distribution. 
Direct measurement of the polarized strange sea density is crucially important 
to understand the flavor structure of the polarized sea quark distribution.

In semi-inclusive DIS, we have an additional ambiguity coming from the 
fragmentation function.
However, the ambiguity can be neglected in the $x$ distribution
of asymmetry, since the kinematical variable related to fragmentation 
is integrated out in this distribution.
The $x$ differential asymmetry is thus sensitive to the behavior of the 
polarized PDFs.
The asymmetries are actually dominated by the contribution from diagrams 
concerned by the strange quark.
However, the predominant gluonic contribution presented in Fig. \ref{diagram} 
(d) is sizable in whole $x$ regions.
Unfortunately, it is not enough to separate the strange quark distribution 
from the gluon distribution in this analysis.
The separation of the strange quark and gluon distributions in CC charm 
production have been discussed by several people 
\cite{Kretzer99-2, Gladilin01}.
They have insisted the possibility which it can be separated by introducing 
some kinematical cuts.
On the other hand, it is expected that the polarized gluon distribution can be 
precisely determined by a measurement of the prompt photon production in 
polarized proton collisions at RHIC \cite{Frixione00} in near future.
It suggests that the polarized strange quark density is effectively extracted 
through the neutrino-induced semi-inclusive $D$/$\Dbar$ production, if the 
polarized gluon distribution is well determined by RHIC experiments.

\section{Conclusions}
Semi-inclusive $D$/$\Dbar$ meson productions in CC DIS in 
neutrino-polarized proton scattering are discussed.
We indicated that the charm quark production in CC DIS is effective to 
directly access the strange quark distribution $s(x, Q^2)$/$\Delta s(x, Q^2)$ 
in the nucleon.
Our knowledge about the polarized sea quark distribution is still deficient 
at present.
Direct measurement of the polarized strange sea density is crucial to 
understand the flavor structure of the polarized PDFs 
and gives us a clue to resolve the so-called proton spin puzzle.
Neutrino experiments using the polarized target which we have discussed here 
might be performed at a future neutrino factory.

The cross sections and the spin asymmetries have been calculated including 
${\cal O}(\alpha_{s})$ NLO corrections with various parametrization models of 
the polarized PDFs.
The prametrization model dependence on the asymmetry is quite large and the 
behavior for the LSS parametrization is especially different among other 
parametrizations. 
Therefore, we can sufficiently test the parametrization model of the polarized 
PDFs in this reaction.
In addition, we have shown that even an ambiguity for the unpolarized PDFs is 
measurable.
In any case the $\Dbar$ production is promising to extract the strange sea 
quark density. 
This is not the case for the $D$ production because of the large valence 
$d_v$ quark contribution over the strange quark contribution.
If the gluon polarization $\Delta g(x, Q^2)$ is fixed by RHIC experiments 
with high accuracy, we can directly extract the polarized strange quark 
distribution $\Delta s(x, Q^2)$.

\end{document}